\documentclass{article}
\usepackage{anysize}
\marginsize{1.5cm}{1.5cm}{1cm}{1cm}
\usepackage{graphicx}
\expandafter\let\csname equation*\endcsname\relax
\expandafter\let\csname endequation*\endcsname\relax
\usepackage{amsmath}
\usepackage{float}
\usepackage{amsfonts}
\usepackage{epstopdf}
\usepackage{amssymb}

\begin{document}

%\affiliation{Departamento de F\'isica Te\'orica de la Materia
%Condensada and Condensed Matter Physics Center (IFIMAC), Universidad Aut\'onoma de Madrid, E-28049 Madrid,
%Spain}
%\email{fj.garcia@uam.es}

\begin{center}

\Large{\textbf{Non-Markovian effects in waveguide-mediated entanglement.}\\}
\vspace{0.5cm}

\large{C Gonzalez-Ballestero, F J Garc\'ia-Vidal, Esteban Moreno}

\vspace{0.3cm}
\normalsize \textit {Departamento de F\'isica Te\'orica de la Materia Condensada and Condensed Matter Physics Center (IFIMAC), Universidad Aut\'onoma de Madrid, E-28049 Madrid,Spain}

\end{center}

\begin{abstract}
We study the generation and evolution of entanglement between two qubits coupled through one-dimensional waveguide modes. By using a complete quantum electrodynamical formalism we go beyond the Markovian approximation. The diagonalization of the hamiltonian is carried out, and a set of quasi-localized eigenstates is found. We show that when the qubit-waveguide coupling is increased, the Markov approximation is not anymore valid, and the generation of entanglement is worsened. 
\end{abstract}

\vspace{2pc}


\section{Introduction.}

During the last years, many efforts have been undertaken to control the interaction between quantum emitters and the electromagnetic (EM) field. One of the principal motivations that drive the interest in this research area lies in quantum information. The development of efficient quantum information devices requires the control of light-matter interaction at a quantum level and, in particular, the precise handling of the entanglement resource. Many promising applications of entangled light-matter states have been proposed, such as quantum teleportation, quantum cryptography or simple quantum logical gates \cite{SINGLEphotonTRANSISTOR,Qgate,QgateEXP}. The fundamental system all these applications rely on is the two-qubit ensemble. Specifically, much interest is focused on generating entanglement by means of coupling the qubits to a common EM environment. In order to pursue this objective, the properties of the surrounding EM modes are crucial. As the free space interaction between emitters is not strong enough for entanglement purposes, additional structures have to be included. These structures modify the density of EM states through the Purcell effect, giving rise to collective phenomena. Two main ways have been proposed to achieve this modifications: optical cavities \cite{CavityREVIEW,DONUTcavity}, in which one or more discrete modes are responsible for the dynamics, and guided modes \cite{WG1,PCrystalold,SLOTwg}, in which qubits are coupled to an EM continuum. Hybrid systems combining both cavities and waveguides have been proposed as well \cite{HYBRIDprb,HybridSUPERCON,HybridNATURE}. Much theoretical and experimental work has been carried out to improve light-matter coupling in various systems, such as photonic crystal cavities \cite{PCcavity1,PCcavity2}, photonic crystal waveguides \cite{PCwaveguide}, and dielectric slot \cite{SLOTwg} and plasmonic waveguides \cite{QuantumOpticsSPP,GenerationSINGLEspp,PRBDiego}. Despite the fact that cavity systems have been studied in more detail, waveguides are also good candidates for a large scale implementation of quantum devices. The feasibility of experimental realization and the possibility of generating long-lived entanglement \cite{PRBDiego,PRLDiego,PhasGate} strongly support this idea. Moreover, waveguides offer an easy way to couple not only qubits but also photons themselves \cite{COUPLEDwaveguidesPHOTONentanglement,EXPERIMphotonENTANGLEMENT}, thus providing an excellent workspace for quantum computation applications.
 
 As commented above, an intense light-matter coupling is extensively seeked by researchers. However, the assumption that the entanglement properties will be enhanced for greater qubit-photon coupling is not always correct, as we demonstrate in this work. Previous investigations have reported the possibility of entanglement generation between qubits coupled to plasmonic waveguides \cite{PRBDiego,PRLDiego,BarangerNONmarkov}. These works made use of the master equation formalism, in which the EM degrees of freedom are traced out. After a so-called \textit{Markov approximation} \cite{FicekLONG,Louisell}, the system under study is described by the density matrix of the two qubit subspace. The dynamics of this density matrix is determined by a master equation through a coherent and a dissipative term. Both contributions contain the effect of the traced EM modes which mediate the interaction. Although very successful in three dimensional (3D) quantum optics calculations, this method has some limitations when the dimensionality of the system is lower. Specifically, if photons are confined in 1D waveguides they undergo successive reflections between the qubits. This, as we will show, can play a key role in the system dynamics, so that averaging the effect of the EM field in this manner is not always a valid approximation. In this paper, instead of the master equation, we solve the hamiltonian completely. The real space formalism we use has been developed by other authors \cite{FanatomCAVITY}, and is widely used in photon scattering problems \cite{SINGLEatomFAN,BarangerMulti,BarangerNONmarkov}. The set of eigenstates found in these works has successfully described the system for a large variety of initial states. Nevertheless, we will show that for certain initial conditions some extra eigenstates must be added in order to describe the system dynamics. We present a detailed analysis of the qubit populations and the degree of entanglement for different regimes. Our results show that the dynamics of the system is Markovian for weak waveguide-qubit coupling. On the other hand, when the coupling is increased a new non-Markovian regime appears, in which the entanglement generation is worsened. The different behavior of the populations in this strongly coupled regime could be used for studying the strong coupling phenomenon in waveguide QED systems.

 The paper organization is as follows: in section 2 we solve Schr\"odinger's equation for the lossless case. This section shows in an easy way how new localized eigenstates appear. In section 3 we diagonalize the hamiltonian in the presence of losses. Once we have solved Schr\"odinger's equation, we study the evolution of the qubit populations in section 4. Finally, conclusions are presented in section 5.
  
     \begin{figure}[ht] \centering
      \includegraphics[scale=0.25]{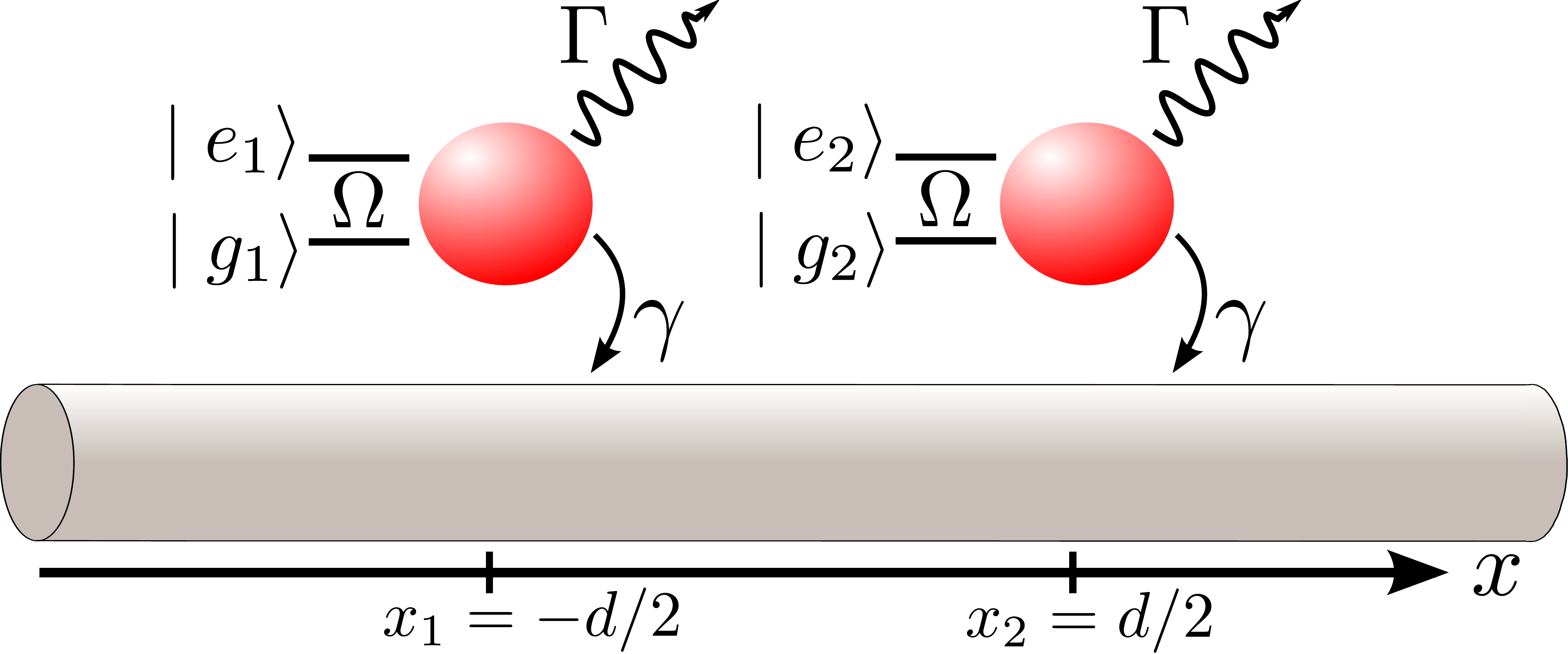} 
      \caption{Two qubits with transition frequency $\Omega$ separated by a distance $d$ are positioned over a waveguide. Interaction between the qubits is mediated by the photonic modes in the waveguide.\label{FIGsystem}}
      \end{figure} 
  
  \section{Lossless system.}
  
  The system under study is composed by two quantum emitters coupled to a one-dimensional infinite waveguide (\ref{FIGsystem}). The waveguide is set along the $x$ axis, and supports a continuum of photonic modes. The dispersion relation, $\omega(k)$, can be approximated linearly within a certain frequency interval \cite{FanatomCAVITY}, $v_g$ being the group velocity. The emitters are modelled as identical two level systems with transition frequency $\Omega$, located at positions $x_{1} = - d/2$ and $x_{2} = d/2$. They interact with the waveguide modes with a coupling energy $\gamma$. We assume that the only coupling between the two emitters is the one mediated by the waveguide modes, hence neglecting the direct coupling energy. This is a realistic assumption for separations in the wavelength scale or greater, as the direct interaction through free space modes decays rapidly with the separation \cite{Novotny}. Losses in the system are taken into account by coupling the qubits to an infinite electromagnetic reservoir with coupling energy $\Gamma$.  In this preliminar section, however, we will assume the system to be completely isolated ($\Gamma = 0$).
  
  The hamiltonian of the lossless system can be expressed in the position basis as \cite{FanatomCAVITY}
   \begin{equation}\label{HamiltonianLR}
    \begin{aligned}
    H_0 = \Omega(\sigma^{\dagger}_{1}
    \sigma_{1} + \sigma^{\dagger}_{2}\sigma_{2})-iv_g\int dx &\bigg( c^{\dagger}_R(x)\frac{\partial}{\partial x}c_R(x)  -c^{\dagger}_L(x)\frac{\partial}{\partial x}c_L(x)\bigg)+
    \\
    +\sum_{j=1,2}V\int dx \delta(x-x_j) \big[ &c^{\dagger}_R(x)\sigma_j + c^{\dagger}_L(x)\sigma_j + h.c.\big],
    \end{aligned}
    \end{equation}
  where we take $\hbar = 1$. Here, $\sigma_{1,2}$ are the lowering operators for the emitter 1 and 2, respectively. The field operators $c^\dagger_{R,L}(x)$ create a right (R) or left (L) propagating photon at the position $x$. In the upper line of equation \ref{HamiltonianLR}, the energy of the emitters and the waveguide modes is contained in the first and second terms, respectively. The second line contains the interaction potential. Note that the coupling constant, $V$, has not dimensions of energy and we have to define the coupling energy \cite{SINGLEatomFANposition} as $\gamma = V^2 / v_g$. The fact that the coupling is not defined as an energy arises naturally when coupling a real continuum to a discrete system \cite{Fanprimero}, as can be checked by taking the continuum limit of the Anderson impurity model \cite{Anderson,Kspace}. Note that the interaction term in  equation \ref{HamiltonianLR} is point-like in space.
  
  Given the symmetry of the system, the natural basis is
  \begin{align} \label{changeBASIS}
  c_{e,o}(x)& = \frac{1}{\sqrt{2}} \left(   c_R(x) \pm c_L(-x)  \right) \\
  \sigma_{e,o} = & \frac{1}{\sqrt{2}} \left(   \sigma_1 \pm \sigma_2  \right),
  \end{align}
  where the indices $e,o$ stand for the even and odd symmetry of the operators. Expressed in this basis, the hamiltonian in equation \ref{HamiltonianLR} splits into uncoupled, independent contributions $H = H_\text{even} + H_\text{odd}$ which can be solved separately. A compact notation for the hamiltonian in both even and odd subspaces is the following (for $j = even, odd$):
   \begin{equation}\label{hamiltJ}
  \begin{aligned}
  H_j = \Omega & \sigma^{\dagger}_{j} \sigma_{j} -iv_g \int dx c^{\dagger}_j(x) \partial_x c_j(x) \\
  + V \int dx \big[\delta(x+d/2) &+\eta_j \delta(x-d/2)\big] \left(c^{\dagger}_j(x)\sigma_j + \sigma^{\dagger}_j c_j(x)\right),
  \end{aligned}
  \end{equation}
  where we have introduced the variable $\eta_{e} = 1 \; ; \eta_{o} = - 1$.

  Along this paper, we will only work in the one-excitation subspace. The diagonalization is carried out by assuming the most generic form for a one excitation Fock state \cite{BarangerNONmarkov} in the subspace $j \; (=even,odd)$, with energy $\epsilon$:
  \begin{equation}\label{FockState}
  \vert \epsilon_j\rangle = \int dx \phi_j(x)c^{\dagger}_j(x) \vert 0 \rangle + \alpha_j \sigma^{\dagger}_j \vert 0 \rangle.
  \end{equation}
  The qubit coefficient $\alpha_j$ and the photon wavefunction $\phi_j(x)$ are the unknowns to be determined. By solving the time-independent Schr\"odinger's equation \cite{BarangerMulti} we find they are related through
  \begin{align}
  (\epsilon - \Omega)\alpha_j & = V(\phi_j(-d/2) + \eta_j \phi_j(d/2)) \label{algebraicRAW1} \\
  (\epsilon +iv_g \partial_x)\phi_j & = V\alpha_j \left[\delta(x+d/2) + \eta_j\delta(x-d/2)\right] .\label{algebraicRAW2}
  \end{align}
  Equation \ref{algebraicRAW2} shows that the photonic wavefunction must be a piecewise continuous plane wave. Then, without loss of generality we can take the next Ansatz for this function:
  \begin{equation}\label{gral ansatz PHI}
  \phi_{j}(x) = e^{i \epsilon x / v_g}
  \Biggm\lbrace \begin{array}{l c r}
  A_j(\epsilon) & \text{for} & x < -d/2
  \\
  t_{0,j} (\epsilon) & \text{for} & -d/2<x<d/2
  \\
  t_{1,j}(\epsilon) & \text{for} & x > d/2.
  \end{array}
  \end{equation}
  By inserting this Ansatz into equations \ref{algebraicRAW1} and \ref{algebraicRAW2} we arrive at a final set of algebraic equations:
  \begin{equation}\label{algebraicFINAL}
  \begin{array}{c}
  (\epsilon - \Omega)\alpha_j = \frac{V}{2}\left(\eta_j(t_{1,j}+t_{0,j})e^{i\epsilon d/2v_g}  + (A_j+t_{0,j})e^{-i\epsilon d/2v_g}   \right)
  \\
  iv_g(t_{1,j}-t_{0,j})e^{i\epsilon d/2v_g} = V\alpha_j
  \\
  iv_g(t_{0,j}-A_j)e^{-i\epsilon d/2v_g} = \eta_j V\alpha_j.
  \end{array}
  \end{equation}
  In this indeterminate system, one of the coefficients (let's say $A_j(\epsilon)$) acts as a quantum number to label different sets of states. Hence, we will have the following cases depending on its value: 
  
  \subsection{Scattering Eigenstates, $A_j(\epsilon) \ne 0$.}
  
\begin{figure}[ht] \centering
   \includegraphics[scale=0.21]{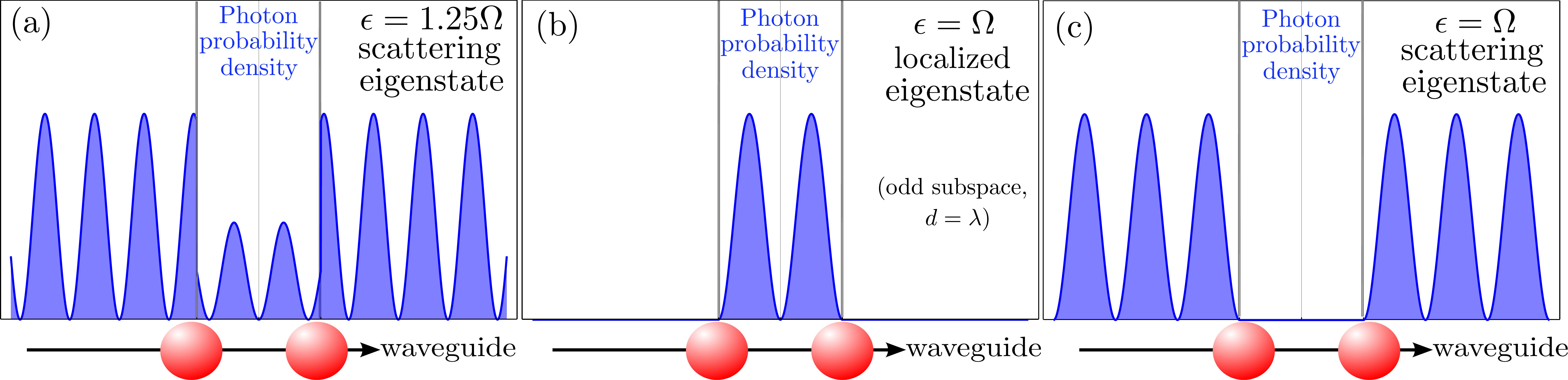}
   \caption{ Photonic probability density for different eigenstates. (a) Scattering, off-resonant eigenstate. (b) Localized state, existing as a stationary wave only in the inter-qubit region. (c) Scattering state in resonance with the emitters frequency $\Omega$. This state is totally reflected.\label{FIGEigenstates}}
   \end{figure}

  If we consider $A_j$ different from zero, we normalize the other three unknowns to $A_j$ to obtain a branch of eigenstates, which we shall call \textit{scattering branch}, with the following coefficients:
  \begin{equation}\label{t0coefficients}
  t_{0,j} = \frac{\epsilon - \Omega}{\epsilon - \Omega + i\gamma\left(1+\eta_je^{i\epsilon d / v_g}\right)}
  \end{equation}
  \begin{equation}\label{t1coefficients}
  t_{1,j} = \frac{\epsilon - \Omega - i\gamma\left(1+\eta_je^{-i\epsilon d / v_g}\right)}{\epsilon - \Omega + i\gamma\left(1+\eta_je^{i\epsilon d / v_g}\right)}  
  \end{equation}
  \begin{equation}\label{ALFAcoefficients}
  \alpha_j = V \frac{e^{-\epsilon d /2 v_g}+\eta_je^{\epsilon d /2 v_g}}{\epsilon - \Omega + i\gamma\left(1+\eta_je^{i\epsilon d / v_g}\right)}.
  \end{equation}
  An example of a scattering eigenstate is shown in figure \ref{FIGEigenstates}(a). This branch of eigenstates has been obtained in previous works for one \cite{PCrystalold,SINGLEatomFAN} and two emitters \cite{BarangerMulti}, and it is sufficient to completely describe any scattering state. However, for other initial configurations, additional eigenstates are needed.
  
  \subsection{Localized Eigenstate, $A_j(\epsilon) = 0$.}

  If we set $A_j$ equal to zero, we find that some states can appear under certain combination of parameters. More precisely, these states only exist when the separation between qubits is equal to half an integer multiple of the emitter characteristic wavelength, i.e., $d = n\lambda/2$ where $\lambda = v_g /2 \pi \Omega$. We will name these the \textit{resonant separations} ($d_\text{res}$) from now on. As a consequence of this condition, the energy of these states has to be $\epsilon = \Omega$ in order to have a nontrivial solution. Moreover, the norm has to be finite so the transmission coefficient must be $t_{1,j}(\epsilon) = 0$.
  
  Provided that the conditions $\lbrace d=d_\text{res};\epsilon = \Omega\rbrace$ are satisfied, system \ref{algebraicFINAL} can be shown to have an additional solution. Only the coefficients $t_{0,j}$ and $\alpha_j$ are different from zero, and are given by
  \begin{equation}\label{coeffsBALLESTERO}
  t_{0,j} = -i\eta_j\frac{V \alpha_j}{v_g} e^{i\pi (d/\lambda)}.
  \end{equation}
  These states are localized in the region between emitters, as shown in figure \ref{FIGEigenstates}(b). The appearance of these states can be understood by noticing that, when a photon with energy $\Omega$ is scattered by a single emitter, the transmission probability is zero \cite{SINGLEphotonSPECTROSCOPY}. Hence, the regions outside and inside the two qubits become totally uncoupled, and incident photons with this energy are completely reflected \cite{PerfectREFLECTIONdipole,SINGLEphotonTRANSISTOR}. This is shown in figure \ref{FIGEigenstates}(c). But the existence of a photon with energy $\Omega$ in the inter-qubit region is also possible. In order for this state to be stationary (and thus an eigenstate of the system) the stationary wave condition must be fulfilled for this qubit-qubit \textit{cavity}. This requirement is obviously satisfied if the separation is half an integer multiple of the wavelength, i.e., $d = d_\text{res} = n\lambda/2$.

  Figure \ref{FIGEigenstates}(b) shows the photon probability distribution of this new localized eigenstate, which is formed by a qubit part and a resonant photonic contribution that exists only inside the cavity. Note that the parameters for which this state appears ($d=d_\text{res}$, $\epsilon = \Omega$) are precisely the zeros in the denominator of the scattering coefficients. Note also that for even resonant separations ($d = 2n\lambda$), the localized state only appears in the odd subspace, and for odd resonant separations ($d = (2n-1)\lambda/2$) it appears only in the even subspace.
  
  The appearance of this finite norm state is essential in order to have a complete basis. Previous works \cite{BarangerMulti,BarangerNONmarkov,SPP2qubits} have studied the scattering of photons with the two emitter system. These new localized states do not play any role in the dynamics of these scattering problems, as their overlap with any incident wavepacket is zero. This is a direct consequence of the condition $A_j(\epsilon) = 0$. However, if the initial state contains some emitter contribution, the addition of these states is essential. Note that, when the separation is not resonant, there is no localized state. Hence, for $d \neq d_\text{res}$ the scattering basis is complete to describe the evolution of the system for any initial conditions. On the other hand, when the separation is resonant the new state appears and has to be taken into account. The importance of the localized state in the time evolution can be quantified in the following way: let us set a resonant separation between qubits, so that a localized state appears in either the even or the odd subspace. We now take an initial state $\vert \Phi (t=0) \rangle$ and define $P_0 = \langle \Phi(t=0) \vert \sigma^{\dagger}_j \vert 0\rangle$. This quantity represents the qubit part of the state in the resonant subspace. This overlap can be expressed as $P_0 = P_\text{sc} + P_\text{loc}$, in which scattering and localized state contributions appear separately. By performing this separation we may quantify both of them. As the localized eigenstate is orthogonal to any scattering one, the fraction $P_\text{loc}/P_{0}$ is constant with time. Its square modulus represents the amount of initial qubit population contained in the localized state. It can be shown to be
  \begin{equation} \label{infinitePOP}
  \bigg \vert \frac{P_\text{loc}}{P_0} \bigg \vert = \frac{1}{1+2\pi(\gamma/\Omega)(d/\lambda)}.
  \end{equation}
  In this simple expression it can be seen that the lower the coupling and the resonant distance, the greater is the importance of this localized state. As an example, for high couplings and low resonant separations, e.g. $\gamma = 0.01\Omega;d=\lambda/2$ the localized state can carry out more than $95\% $ of the probability, thus governing the dynamics in the resonant subspace.
  
  Now that the localized states are well understood, we can go one step beyond and take losses into account. After solving the more realistic lossy hamiltonian, we will observe that the effective coupling to additional decay channels produces the localized state to become not a single, discrete state but a whole new continuous branch.
  
 \section{Lossy system.}
 
 In this section we solve the system with losses by adding to the qubits a finite decay rate into the EM environment. Losses in the waveguide are not considered, although a brief comment on finite photon propagation lengths can be found at the end of this paper. The problem of introducing losses in this system has already been studied \cite{FanatomCAVITY}. Previous works \cite{BarangerMulti,2photonsFAN} introduce losses by adding an imaginary part to the qubit frequency, i.e., $\Omega \mapsto \Omega - i\Gamma/2$. This modification, as we will see below, is enough to describe scattering problems. However, for certain initial conditions some extra states are needed, as we have already seen in the lossless situation. Moreover, when adding an imaginary part to the qubit frequency, the hamiltonian becomes non hermitian. Even though it can be formally diagonalized to obtain a modified scattering branch of eigenstates, we are not able to recover the localized states when we take the limit $\Gamma \to 0$. Additionally, it can be shown that the closure relation formed by the eigenstates of this non hermitian hamiltonian is not complete for particular initial states. This indicates that there must be missing eigenstates. For all these reasons, a more complete description of the losses is necessary.
 
 Our model will be similar to that used in the previous section, the hamiltonian being
 \begin{equation}\label{HlossyTOT}
 H = H_{0} + H_{\text{r}} + H_{\text{c}},
 \end{equation}
 where $H_0$ is the original lossless hamiltonian (equation \ref{HamiltonianLR}). The second term, $H_{\text{r}}$, is the contribution of the reservoir modes,
 \begin{equation}\label{Hlosses}
 H_{\text{r}} = -iv\int dz P^{\dagger}_1(z) \partial_z P_1(z) -iv\int dz P^{\dagger}_2(z) \partial_z P_2(z).
 \end{equation}
 Here, we have modelled the free space EM environment of each emitter as one infinite continuum of photonic modes. Equation \ref{Hlosses} is similar to the hamiltonian of two infinite waveguides (second line in equation \ref{HamiltonianLR}). In this case, however, they would be \textit{unidirectional} waveguides, in the sense that they do not have two but a single propagation direction. This keeps their main effect on the qubit populations providing an exponential decay, and simplifies the calculations. These 1D reservoirs are independent of each other and are set along the $z$ axis. They are characterized by the creation operators $P_1^{\dagger} (z)$ and $P_2^{\dagger}(z)$ respectively, and a group velocity $v$ which we will consider to be $v = v_g$ for simplicity. The third term in equation \ref{HlossyTOT}, $H_{\text{c}}$, represents the coupling of the reservoir with the qubit states. This contribution is expressed as
 \begin{equation}\label{Hlosscoupling}
 H_{\text{c}} = K\int dz \delta(z) \bigg[  P^{\dagger}_1(z) \sigma_1 + P^{\dagger}_2(z) \sigma_2 + h.c. \bigg],
 \end{equation}
 where $K$ is the coupling strength. The coupling energy is then defined as $\Gamma = K^2 / v$, as discussed in previous section. The interaction is set to take place in $z = 0$ for both reservoirs.
 
 Once we have introduced a decay rate to the guided modes, $\gamma$, and a second one to free space modes, $\Gamma$, we are in a more realistic situation. Here, we can define the usual figures of merit in this kind of problems, i.e., the Purcell factor $F_P$ and the $\beta$ factor of a single emitter \cite{Novotny}. The first one is defined as the enhancement of the decay rate with respect to the free space case, $F_P = (2\gamma+\Gamma)/\Gamma$. The $\beta$ factor is defined as the amount of radiation coupled to guided modes, i.e., 
 \begin{equation}\label{beta}
 \beta = \frac{2\gamma}{2\gamma + \Gamma}.
 \end{equation}
 The factor 2 appears because, as the waveguide has two propagation directions, the total decay rate to the guided modes is $2\gamma$. Reservoir states, however, are unidirectional as mentioned above, hence being $\Gamma$ the total decay rate. 
 
  In order to diagonalize hamiltonian \ref{HlossyTOT}, let us express the reservoir operators in the even/odd basis as
 \begin{equation} \label{lossEO}
 P^{\dagger}_{e,o} (z) =\frac{1}{\sqrt{2}} \left( P^{\dagger}_1 (z) \pm P^{\dagger}_2 (z) \right).
 \end{equation}
 Under this change of basis, both $H_{\text{r}}$ and $H_{\text{c}}$ split into independent even and odd contributions:
  \begin{equation}\label{HlossDEF}
  \big(H_{\text{r}} + H_{\text{c}}\big)_j = -iv\int dz P^{\dagger}_j(z) \partial_z P_j(z) + K\int dz \delta(z) \big[ P^{\dagger}_j(z) \sigma_j +  h.c. \big],
  \end{equation}
 for $j = even, odd$. The expression of the one-excitation Fock state is now
 \begin{equation}\label{FockSTlossy}
 \vert \epsilon_j \rangle = \bigg(\int dx \phi_j(x) c^{\dagger}_j(x) + \int dz \psi_j(z) P^{\dagger}_j(z) + \alpha_j \sigma^{\dagger}_j \bigg) \vert 0 \rangle,
 \end{equation} 
 where the Ansatz for $\phi_j (x)$ is the same as in the lossless case (equation \ref{gral ansatz PHI}) and the Ansatz for the photonic wavefunction of the reservoir, $\psi_j(z)$, is
 \begin{equation}\label{AnsatzPSIgen}
 \psi_j(z) = e^{i \epsilon z /v_g} \Big\lbrace \begin{array}{cl l} a_j & \text{for} &  z< 0 \\ b_j & \text{for} & z>0.\end{array}
 \end{equation}
 We can diagonalize now, following the same steps as in previous section. In this case, the outcome of this diagonalization is a system of $4$ algebraic equations with $6$ unknowns. We choose the independent variables to be $A_j(\epsilon)$ and $a_j(\epsilon)$, so that the remaining variables are determined if these two are known. It can be demonstrated that no eigenstate exists with both coefficients equal to zero, and hence only two possibilities remain:
 
 \subsection{Scattering branch, $a_j = 0, A_j \ne 0$.}
 
 For these values of the parameters, all unknowns can be normalized to $A_j$, so that we obtain a $4  \times 4$ algebraic system. The solution for the reservoir can be shown to be
 \begin{equation}\label{bSCAT}
 b_j = -i\frac{K}{v_g}\alpha_j.
 \end{equation}
 The solutions for the coefficients $\lbrace t_{0,j},t_{1,j},\alpha_j\rbrace$ are identical to those of the lossless case (equations \ref{t0coefficients}, \ref{t1coefficients}, \ref{ALFAcoefficients}), with an additional imaginary part in the qubit frequency. That is, by performing the substitution $\Omega \mapsto \Omega - i\Gamma/2$, previous works obtain exactly this branch of eigenstates. The scattering branch is thus composed of correct eigenstates. However, as we have already remarked above, the subspace spanned by them is not always enough to describe the state of the system. 
 
 \subsection{Quasi-localized branch, $A_j = 0, a_j \ne 0$.}

 Let us explore the second possible parameter combination and look for the equivalent of the localized state we found in the lossless case. When taking $A_j = 0$, as we still have an arbitrary parameter, a solution of the Schr\"odinger equation exists for any energy. That is, there are no constraints over the parameters. Consequently, we will have a second continuum of states instead of the discrete one obtained in the lossless case. The expressions of the coefficients, normalized to $a_j$, are
 \begin{equation} \label{LOCbranch2}
 t_{0,j} = -i\frac{\sqrt{\gamma \Gamma}  e^{i \epsilon d/2v_g} }{\epsilon - \Omega +i\frac{\Gamma}{2} +i\gamma \left(1+\eta_j e^{i\epsilon d/v_g}\right)} 
 \end{equation}
 \begin{equation}\label{LOCbranch3}
 t_{1,j} = -i\frac{\sqrt{\gamma \Gamma}  (e^{i \epsilon d/2v_g}+\eta_j e^{-i \epsilon d/2v_g}) }{\epsilon - \Omega +i\frac{\Gamma}{2} +i\gamma \left(1+\eta_j e^{i\epsilon d/v_g}\right)}
 \end{equation}
 \begin{equation}\label{LOCbranchALFA}
 \alpha_j = \frac{K}{\epsilon - \Omega +i\frac{\Gamma}{2} +i\gamma \left(1+\eta_j e^{i \epsilon d/v_g}\right)}
 \end{equation}
 \begin{equation} \label{LOCbranch1}
 b_j = 1-i\frac{K}{v}\alpha_j.
 \end{equation}
 
\begin{figure}[ht] \centering
     \includegraphics[scale=0.3]{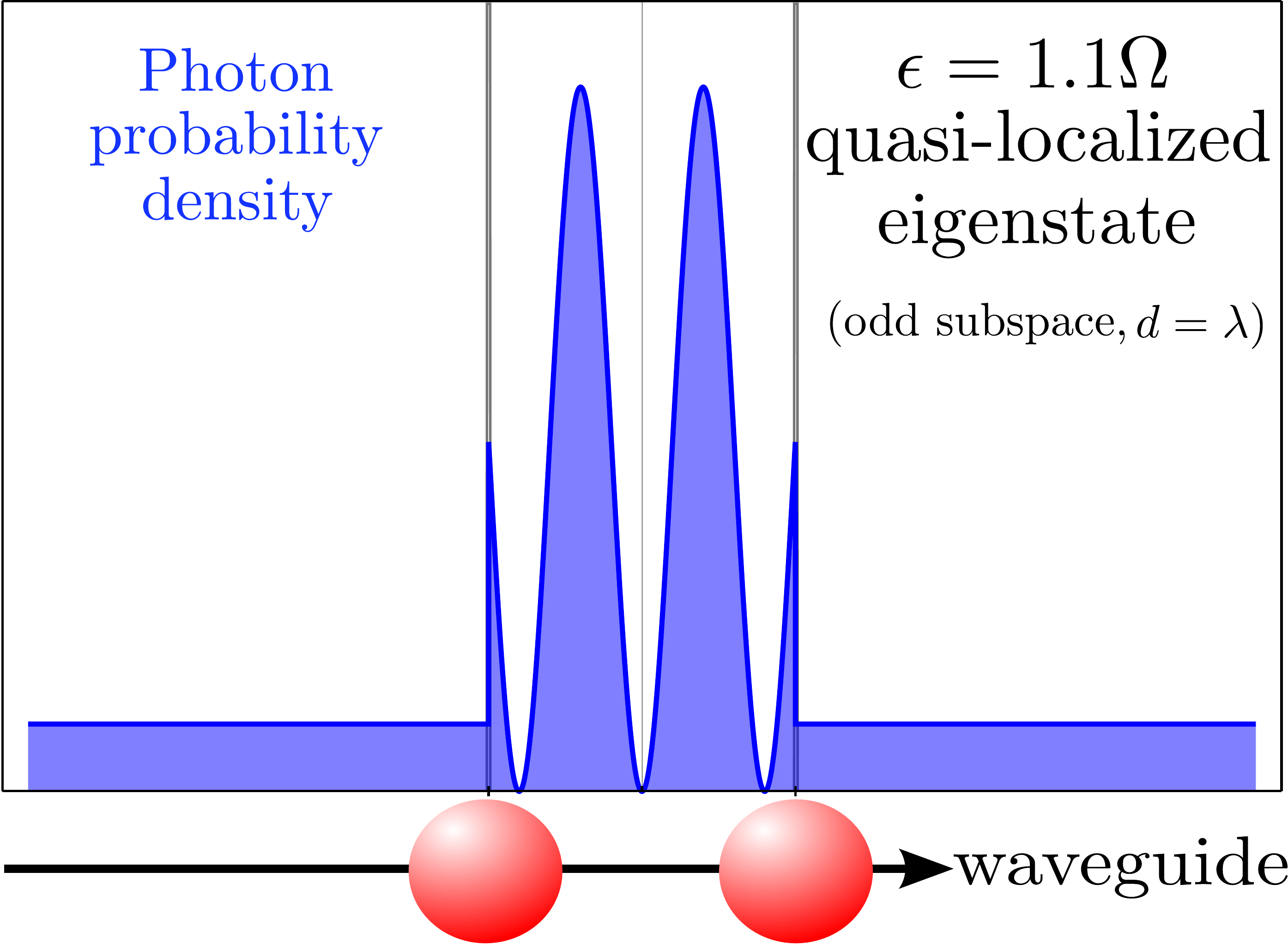} 
     \caption{Photon position probability density for the lossy hamiltonian. As the energy $\epsilon = \Omega$ is not in resonance with the inter-qubit separation, photons leak out of the qubit-qubit cavity.\label{LossyEIGENSTATE}}
     \end{figure}

 It can be demonstrated that the subspace spanned by this branch is orthogonal to the scattering subspace. Another important feature is that, as we take the limit $\Gamma \to 0$ (i.e., $ K \to 0$), we have $t_{1,j}\to 0$ and a factor $\delta(\epsilon - \Omega) \delta(d - d_\text{res})$ appears in both $\alpha$ and $t_{0,j}$. Hence, we recover the single localized state we got in the lossless case. Note that, for any qubit separation, we can have $t_{1,j}=0$ at a given energy (such that the numerator in equation \ref{LOCbranch3} cancels out), even for $\Gamma \ne 0$. In other words, for a given separation $d$ there is always an eigenstate whose wavelength is in resonance with the cavity. That eigenstate exists only in the inter-qubit region, so that it is fully localized and decays only to the reservoir. Note that, as equation \ref{LOCbranchALFA} shows, the maximum qubit occupation is achieved for $\epsilon = \Omega$. Thus, the situation in which the resonant, localized eigenstate is $\epsilon = \Omega$ will be the most interesting for entanglement purposes. This will be achieved precisely for $d = d_\text{res}$. For off-resonant eigenstates,  according to equation \ref{LOCbranch3}, a certain amount of probability is leaking outwards. Figure \ref{LossyEIGENSTATE} shows an example of these quasi-localized eigenstates.
 
  By using the expression of the Fock state (equation \ref{FockSTlossy}) it is also possible to check that the overlap of any eigenstate in the quasi-localized branch with any incident, pure photonic wavepacket in the waveguide is zero. This makes the scattering branch to be complete for any scattering problem. However, if the initial state contains a non-zero qubit component, the localized states play a key role in the dynamics of the system. As we will see below, for resonant qubit separations a single excited qubit decays with a rate proportional to $2\gamma$ due to the scattering branch contribution, whereas the localized branch adds a typical decay $\Gamma$. The competition between these two times and the photon travel time between the qubits $t =d/v_g$ gives rise to various phenomena.

 \section{Populations and entanglement generation.}
 
 The time evolution of qubit populations has been already studied in previous works \cite{PRBDiego,PRLDiego}. Specifically, they have focused in the possibility of entanglement generation. That is, the spontaneous evolution of an initially unentangled system into an entangled state. It is particularly interesting to study the evolution of the state $\vert \Phi (t=0) \rangle =\sigma^{\dagger}_1 \vert 0 \rangle$, i.e., first emitter in the excited state. By using a master equation formalism, the above mentioned works have demonstrated that both symmetric and antisymmetric qubit states decay exponentially. One of them becomes subradiant (decay rate $<2\gamma + \Gamma$) whilst the second one becomes superradiant (decay rate $>2\gamma+\Gamma$). The values of both decay rates depend strongly on the inter-qubit separation. Particularly, when this separation is resonant, the decay rate of the subradiant state into the guided modes becomes zero. As a consequence, its population remains almost constant with time, slowly decaying into the reservoirs with decay rate $\Gamma$. This method provides an efficient way of generating a maximally entangled state. However, the formalism in sections 2 and 3 allows us to go beyond this master equation approach and explore the validity of the Markov approximation.
 
 In this section, we set the same initial state of the system $\vert \Phi (t=0) \rangle =\sigma^{\dagger}_1 \vert 0 \rangle$. As we have a complete set of eigenstates, we can write explicitly the closure relation and obtain the time evolution of this state, $\vert \Phi (t) \rangle$, in the usual way:
 \begin{equation}\label{timeevol}
 \vert \Phi (t) \rangle = \sum_{j=\text{e,o}} \frac{1}{2\pi v_g}\sum_{m=\text{S,Q}}\int d\epsilon e^{-i\epsilon t} \vert \epsilon_{j,m} \rangle \langle \epsilon_{j,m} \vert \Phi(0)\rangle,
 \end{equation}
 where the index $j \; (= even,odd)$ sums over both subspaces, and the index $m$(= Scattering, Quasi-localized) sums over both branches of eigenstates. We are interested in the time-evolution of the populations of the symmetric and antisymmetric states, defined as
 \begin{equation}\label{populations}
 \rho_{\pm\pm}(t) = \langle\pm\vert \Phi (t) \rangle\langle \Phi (t)\vert \pm \rangle,
 \end{equation}
 as well as in determining the degree of entanglement between the qubits.  In order to compare with previous results \cite{PRLDiego}, we will use Wooters concurrence \cite{Wooters} to quantify the entanglement. For the initial state considered the complete expression for the concurrence is reduced to \cite{FicekSHORT}:
 \begin{equation} \label{Wooters}
 C(t) = \frac{1}{2}\sqrt{[\rho_{++}(t) -\rho_{--}(t)]^2+4\text{Im}[\rho_{+-}(t)]^2}, 
 \end{equation}
 where $\rho_{+-}(t) =\langle + \vert \Phi (t) \rangle \langle \Phi (t) \vert - \rangle $ represents the off-diagonal part of the reduced density matrix in the subspace spanned by the pure qubit states $\lbrace \vert g_1 g_2 \rangle,\vert - \rangle = \sigma^{\dagger}_o \vert 0\rangle,$ $\vert + \rangle  = \sigma^{\dagger}_e \vert 0\rangle,\vert e_1 e_2 \rangle \rbrace$. The concurrence and populations arising from our numerical calculations can be compared with the analytical expression obtained within the Markov approximation. The difference between both results will be a sign of non Markovian behavior.
 
 Note that the waveguide-mediated interaction will result in interesting dynamics for both populations and concurrence, as photons can undergo successive reflections inside the inter-qubit cavity. On the other hand, as the emitters are not coupled through the reservoirs, these EM modes will not add interesting features apart from the well known exponential decay. Thus, we will consider only the case $\Gamma < \gamma$, in which the waveguide modes prevail and determine the dynamics. As we are mainly interested on the effects of varying the qubit-waveguide coupling, we will modify the relation $\gamma/\Omega$, keeping the ratio $\Gamma/\gamma = 0.1$ constant along all this work. This corresponds to a Purcell factor $F_P = 21$ ($\beta = 0.95$). Although larger than the typical values in dielectric waveguides \cite{EXPERIMshanhuiFAN,dielectricbeta2}, it lies well below the experimentally observed values for photonic crystal waveguides \cite{HUGHES,PCbeta} ($F_P \lesssim 30$ in the THz regime) and both plasmonic \cite{Purcell1,Purcell2,EXPERIMwaveguideSPP_NVcenter,PhasGate} and slot waveguides \cite{SLOTwg} ($F_P \lesssim 50$).
 
  As a final setting, let the group velocity be $v_g = c$ and study the system for different values of the separation $d$. This does not imply loss of generality, as a reduction in $v_g$ is equivalent to a corresponding increment in the separation $d$: both imply an increase in the time spent by photons to cover the distance between emitters. Different regimes appear in the time evolution, depending on the ratio between the coupling $\gamma$ and the qubit frequency $\Omega$. 
 
   \begin{figure}[ht] \centering
   \includegraphics[scale=0.33]{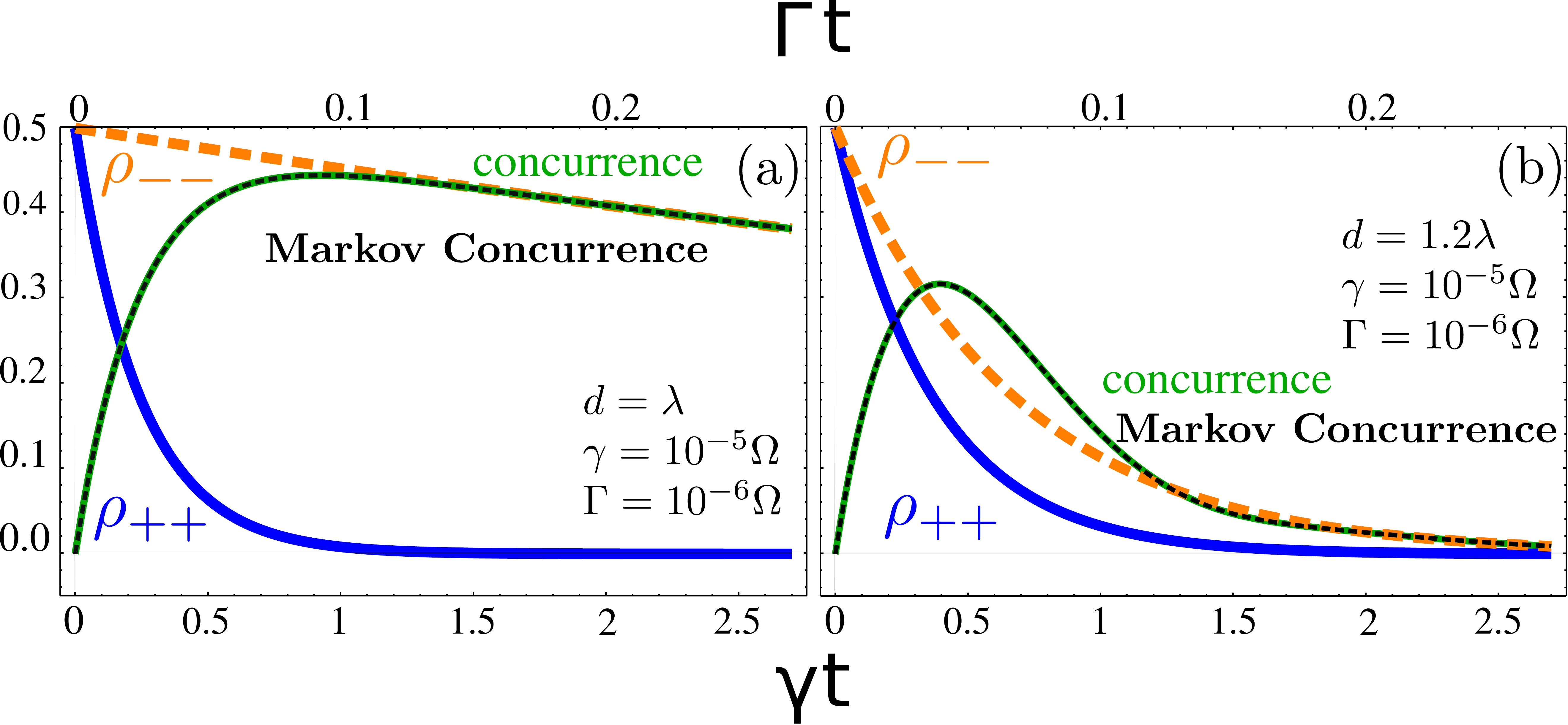}
   \caption{ Time evolution of the populations of the qubit symmetric and antisymmetric states for resonant (a) and non resonant (b) qubit separations. Numerical and analytical Markovian concurrences are also displayed. For weakly coupled qubits, Markov approximation is perfectly valid.\label{FIGmarkovpops}}
   \end{figure}
 
 Markovian results \cite{PRBDiego} are recovered in figure \ref{FIGmarkovpops}. Here, both populations and concurrence are displayed when the coupling is set to be very small, $\gamma << \Omega$. In this situation, the decay time of qubit 1 is much higher than the time spent by photons to reach qubit 2. Hence, we can consider the interaction to be instantaneous and the Markovian approximation is valid. If the qubit separation is resonant [figure \ref{FIGmarkovpops}(a)], the decay of the antisymmetric state $\vert - \rangle$ into the waveguide modes is totally supressed and thus losses are the only decay channel. Note that, for other resonant separations ($d = \lambda/2, 3\lambda/2,...$), the even modes would be uncoupled instead of the odd ones. This effective uncoupling can be understood by looking at the expression of the transmission coefficient (equation \ref{LOCbranch3}). In this low coupling case, the factor $t_1$ is sharply peaked around the qubit frequency, $t_1 \propto (\epsilon-\Omega+i\delta)^{-1}$ where $\delta$ is approximately independent of the energy. This expression is identical to the transmission coefficient for the single emitter case \cite{SINGLEatomFAN}. Consequently, the collective qubit states will behave in a similar manner as single emitters coupled to the waveguide. When the separation is resonant, the coupling for the subradiant state is $\delta \approx 0$. As a consequence, the collective antisymmetric state becomes decoupled from the waveguide. The superradiant state decays rapidly as its coupling is maximum. On the other hand, when the separation deviates slightly from the resonant value, $\delta \ne 0$. Then, the symmetric and antisymmetric states behave in the same way: they are both coupled to the waveguide, as no real energy cancels out the denominator of $t_1$. Hence, they decay exponentially into the waveguide modes as shown in figure \ref{FIGmarkovpops} (b).
 
\begin{figure}[ht] \centering
  \includegraphics[scale=0.33]{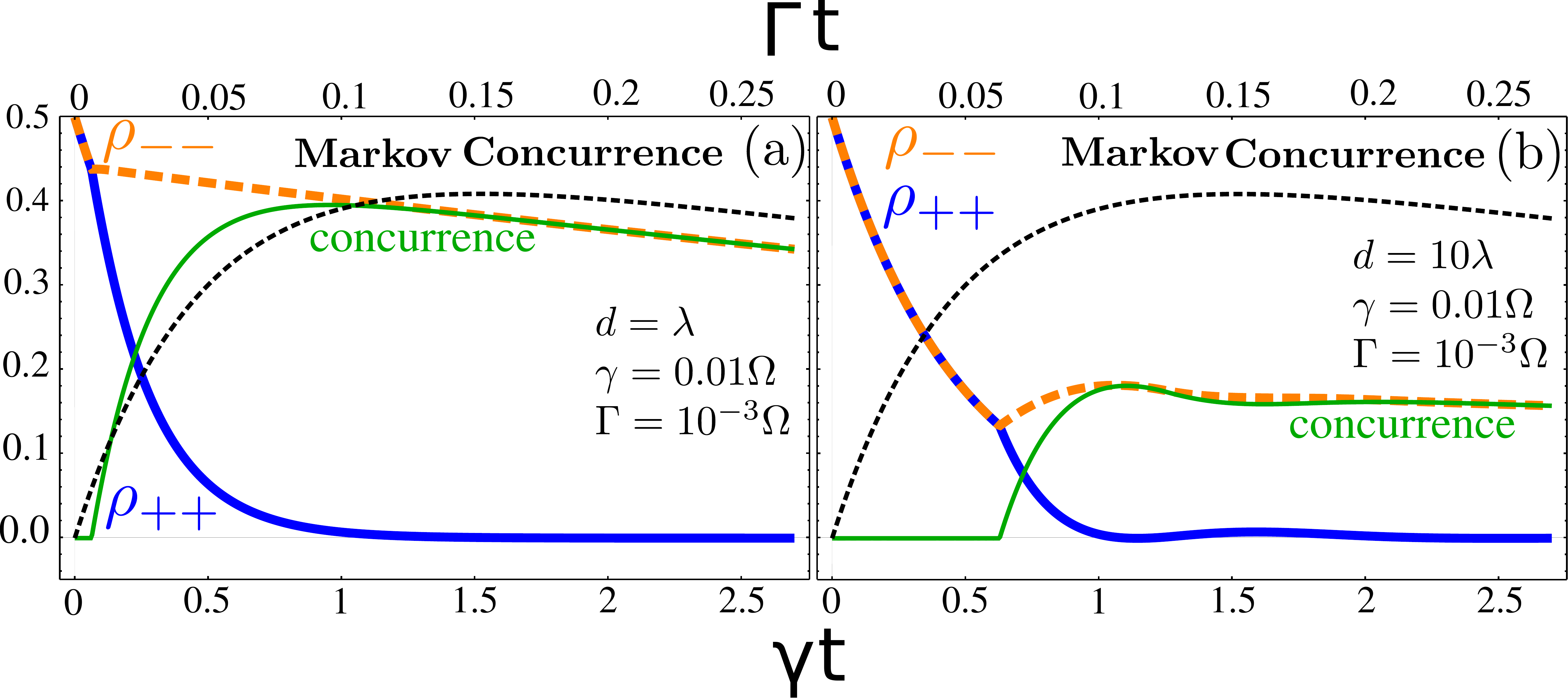}
  \caption{(a) As the qubit-waveguide coupling is increased, both populations and concurrence start to be different from the Markovian case. (b) When the separation is greater we go further outside the Markovian region, as the retarded interaction has to be taken into account.\label{FIGgoingout}}
 \end{figure}

 So far we have recovered the Markovian behavior of the system. However, this collective evolution is modified when the parameter $\gamma$ is different. Specifically, the system is Markovian if the qubit-qubit interaction can be considered to be instantaneous. That is, if the time spent by virtual photons to cover the inter-qubit separation is negligible compared to the qubit decay time $(2\gamma+\Gamma)^{-1}$. If they are comparable, however, non Markovian effects appear and the long time entanglement is destroyed. To explore the regions out of the Markovian regime we can either increase the couping $\gamma$ or the separation between the emitters. As we have previously mentioned, a longer separation between qubits is equivalent to a decrease of the group velocity $v_g$. As a consequence, the coupling $\gamma = V^2/v_g$ is increased in an indirect way. These effects are displayed in figure \ref{FIGgoingout}. Note that we are considering only resonant separations from now on. When the qubit-waveguide coupling is increased  [figure \ref{FIGgoingout}(a)], the lifetime of the excited qubit 1 becomes comparable to the time spent by the emitted photons to reach qubit 2. Thus, both time scales play a role in the system dynamics and the Markov approximation starts to lose validity. Regarding the denominator in equations (\ref{LOCbranch2}, \ref{LOCbranch3}, \ref{LOCbranchALFA}), it cannot be approximated to $\epsilon -\Omega +i\delta$ anymore. Now, the effective coupling $\delta$ depends on the energy and thus the time evolution is modified. Specifically, the exponential term in this denominator, which represents the effect of the second qubit, is not a negligible phase and will be \textit{turned on} at a retarded time. This will modify drastically the system dynamics. In figure \ref{FIGgoingout}(a)  the coupling still fulfills approximately $\gamma << \Omega$ and then deviations from the Markovian situation are small. Note that as we increase the coupling, values for the concurrence become smaller than before. This means that, in opposition to what may have been expected, the higher the coupling, the lower the amount of entanglement and its lifetime.
 
  A completely non Markovian behavior can be observed in figure \ref{FIGgoingout}(b). Here the separation has been chosen as $d = 10\lambda$. As remarked above, this can be seen as an effective increase in the single qubit coupling to the waveguide. We can observe two clearly different regions: for $\gamma t < 2\pi (\gamma/\Omega) (d/\lambda) \approx 0.63 $, the qubit 1 is decaying just as a single emitter coupled to a waveguide \cite{SINGLEatomFANposition}, i.e., an exponential decay with a rate $2\gamma + \Gamma$. This is a consequence of the retarded interaction, which is produced by virtual photons. As they have not yet reached the second qubit, the evolution is exactly the same as that of the single emitter. This can be also understood by the denominator in equations (\ref{LOCbranch2}, \ref{LOCbranch3}, \ref{LOCbranchALFA}): the phase factor $\exp(i\epsilon d/v_g)$ adds poles to our complex plane integral (equation \ref{timeevol}), which represent collective states. However, for short times they are out of our integration contour and do not contribute. Physically this implies that collective states are not taking part in the dynamics as the interaction has not been turned on yet. Hence, both populations $\rho_{++}$ and $\rho_{--}$ decay exponentially with exactly the same behavior. The concurrence remains zero because only qubit 1 is taking part in the evolution, so that no entanglement is possible. On the other hand, for $\gamma t > 2\pi (\gamma/\Omega) (d/\lambda) \approx 0.63$, photons have reached qubit 2, so that it starts becoming involved in the dynamics of the system. Only after this moment the interaction between qubits is turned on and collective effects (and entanglement) appear. In this regime, the even and odd qubit state populations deviate from each other: the symmetric one decays with a higher decay rate as the separation is resonant for the odd modes; the antisymmetric one remains approximately constant, decaying only to the reservoir. In other words, the localized eigenstate $\epsilon = \Omega$ does not decouple from the waveguide until the effect of the second qubit appears. This collective behavior is reminiscent of that predicted within the Markovian regime. However, the delayed interaction changes completely the final values of the populations. As the symmetric state is depopulated for long times, concurrence follows the antisymmetric population in this limit. In the case of figure \ref{FIGgoingout}(b), entanglement generation is clearly worsened with respect to the Markovian case. This, as we have commented, is a direct consequence of the increase in the qubit-waveguide coupling.
 
  Note that, in figure \ref{FIGgoingout}(b), subtle traces of a damped oscillatory behavior can be observed in the populations. These oscillations are caused by multiple reflections of the light pulses inside the inter-qubit \textit{cavity}. Photons corresponding to these pulses are contributions from eigenstates with energy close to resonance, so that they undergo some internal reflections before escaping the inter-qubit cavity. This partial resonance appears because the system is far from the Markovian regime, in which the cavity resonated only for one value of the energy. Here, then, we have a broader resonance spectra for the qubit-qubit cavity. This broadening can be understood by noticing that the total transmittance $\vert t_1 \vert^2$ has approximately a Lorenzian shape, whose width is given by the coupling term. Hence, a higher coupling allows more energies to be close to resonance. For long times, all these quasi-resonant photons abandon the inter-qubit cavity and only the really resonant one ($\epsilon = \Omega$) remains. Consequently, the amplitude of these oscillations in the populations decays as we can see in the figure. This oscillatory or transient region is then an intermediate regime between the single qubit decay regime, in which both emitters can be treated independently, and the region of collective evolution, in which the entangled antisymmetric state evolves as a whole entity. In figure \ref{FIGgoingout}(b), these oscillations have tiny importance in the global dynamics because the coupling is small enough for the collective state to arise in a short time. Nevertheless, in the strong coupled case they will govern the time evolution, as we can see below.  
  
\begin{figure}[ht] \centering
   \includegraphics[scale=0.35]{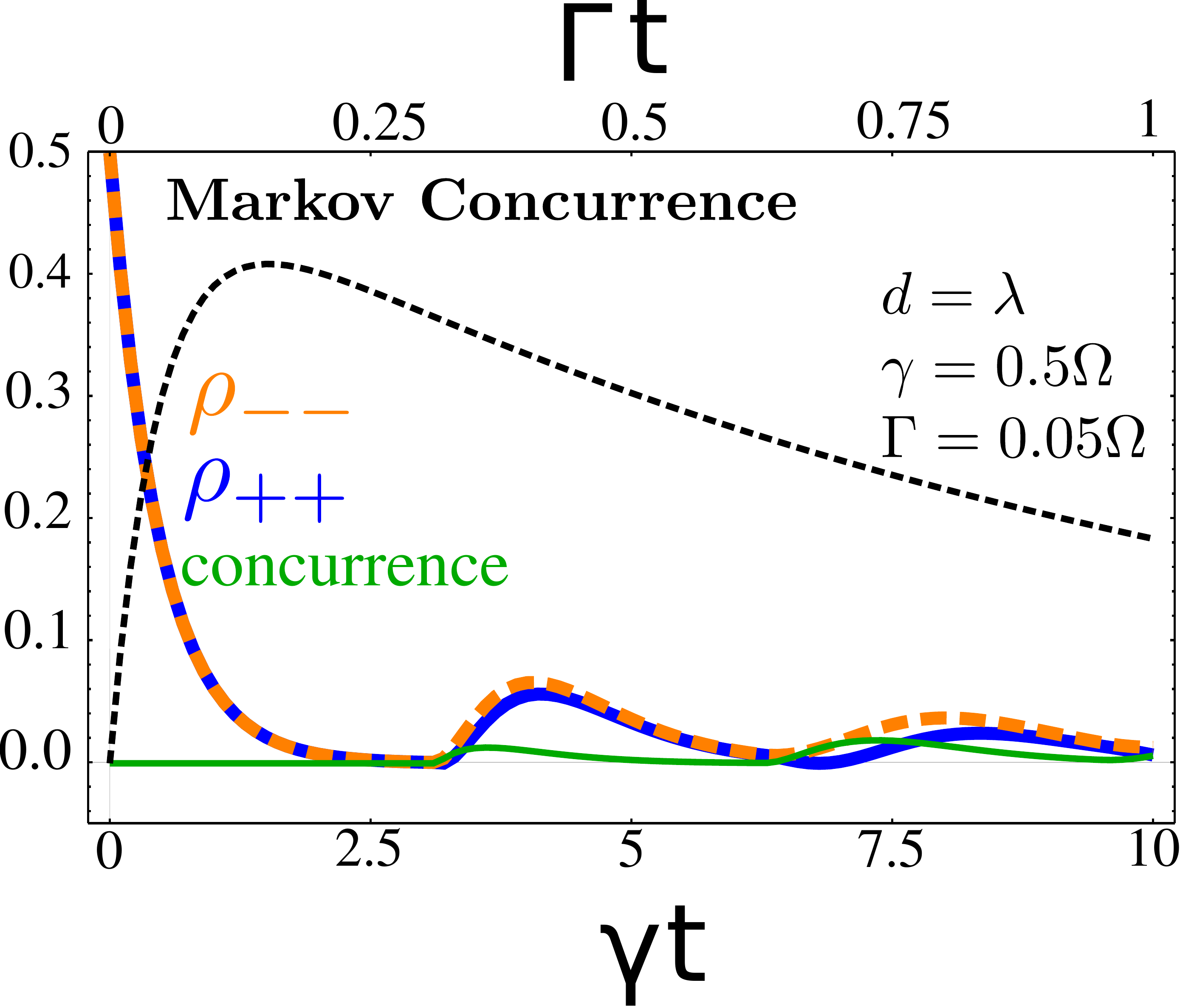}
   \caption{High coupling case: time evolution of the populations of the qubit symmetric and antisymmetric states, as well as the concurrence. In this regime, no significant collective effects appear and the dynamics is described by the single emitter contribution.\label{FIGultrastrong}}
  \end{figure}

  Let us finally explore the case of strong qubit-waveguide coupling in figure \ref{FIGultrastrong}. Here the coupling energy is increased to $\gamma = 0.5\Omega$. This value lies within the ultra-strong coupling regime of cavity QED \cite{ANNPHYSgamma}. Athough values up to $\gamma = 20\Omega$ have been theoretically predicted in particular systems \cite{ANNPHYSgamma,PRBgamma}, this coupling is above the values observed in the experiments \cite{NATUREgamma} ($\gamma \approx 0.12 \Omega$). However, dynamics in this regime shows interesting features that can be observed if the coupling is increased in other ways (e.g. by increasing the number of emitters). In this situation, qubit 1 is so strongly coupled to the photonic modes that it decays completely before the photons reach the second emitter. Hence, as we see in figure \ref{FIGultrastrong}, either the first or the second qubit are populated, but not both at the same time. In other words, the populations of symmetric and antisymmetric qubit states are the same, so that entanglement formation is precluded. Revivals or beats in these populations can be observed, due to succesive reflections of the light pulse inside the cavity. Note that this photonic wavepacket contains a contribution from all energies in a wide range around the resonant one, $\epsilon = \Omega$. It travels inside the cavity for a long time, due to the broader reflectivity spectrum of the qubit. This broadening, as commented above, is a direct consequence of increasing $\gamma$.
  
 It can be observed that both populations are out of sync for  $\gamma t \gtrsim 2\pi$. This is the fingerprint of a change from the single qubit regime to the transient regime commented above. In this transient region, the difference between symmetric and antisymmetric populations will increase slowly. The cause of this is the variation of the shape of the photonic wavepacket inside the qubit-qubit cavity, as the non-resonant photons are slowly escaping. This process takes a long time, as the energies near $\Omega$ are very close to resonance and leak out slowly. At a sufficiently long time, photons of all frequencies will have leaked outside the cavity except the pure resonant one, $\epsilon = \Omega$. Then, the collective regime is achieved, and the system evolves collectively as seen in previous cases: the superradiant state decays quickly to the waveguide modes and the subradiant one uncouples, decaying only to the reservoir. This collective mode has, however, tiny importance in the dynamics for two main reasons: first, as commented in section 2, the importance of the localized states with respect to the scattering ones decreases with $\gamma$ (equation \ref{infinitePOP}). Hence, even in the lossless case the maximum value for the concurrence through all the time evolution would be $C_{\text{lossless}} \simeq 0.025 $ (for the parameters in figure \ref{FIGultrastrong}). Secondly, note that the higher the coupling, the stronger is the effect of the losses. This is a consequence of the fact that the collective regime appears for very long times. Hence, a significant amount of probability has decayed into the reservoirs by this moment. This decreases the maximum value of the concurrence below $0.02$, hence reducing even more the importance of the collective effects. The strong qubit-waveguide coupling thus practically supresses the entanglement formation, the dynamics of the system being mainly described  by the individual emitters.
 
 So far we have studied the time evolution of the qubit populations. Now, let us focus on the evolution of the probability that goes out of the qubit states, i.e., the photon position probability density. As stated before, the formalism used in this work allows us to explore any possible values for the system parameters, even outside the Markovian regime. Another advantage is the possibility of studying the photonic part of the quantum state. Previous studies used the reduced density matrix of the qubit-qubit system, tracing out the electromagnetic degrees of freedom and thus losing all information about them. Here, we can explicitly calculate the photon position probability density, which is equivalent to the emitted photon intensity profile. This is a powerful tool for checking the behavior of the system.
 
   \begin{figure*}[ht] \centering
   \includegraphics[scale=0.2]{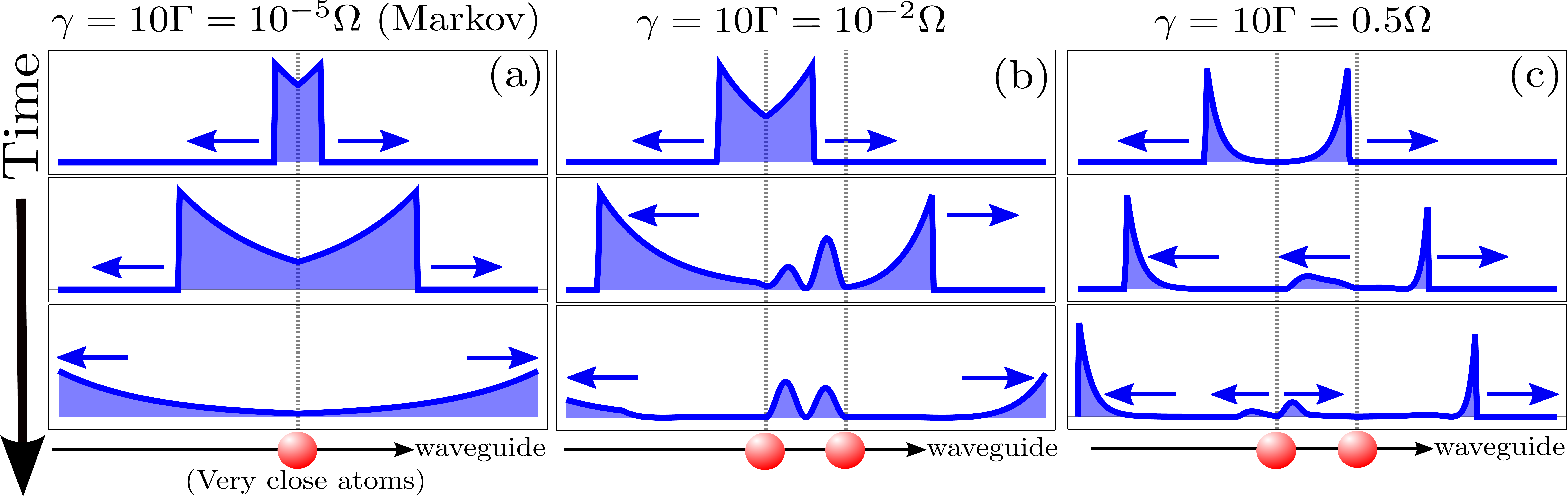}
   \caption{Time evolution of the emitted photon position probability density. (a) Markovian case. Both qubit symmetric and antisymmetric state evolve in the same way as a single emitter. (b) Strongly coupled case. The retarded interaction lowers the probability of having long-time entanglement. (c) Ultra-strong coupled case, in which no entangled collective state arises.\label{FIGframes}}
   \end{figure*} 
 
 Figure \ref{FIGframes} displays the photon probability distribution  for three different cases. First one [figure \ref{FIGframes}(a)] shows the Markovian case, with the same parameters as in figure \ref{FIGmarkovpops}(a). For a better visualization, the position scale in the horizontal axis has been taken to be much larger than the inter-qubit separation $d$. The emission properties of the system in the Markovian limit are very similar to those of a single emitter. This is due to the de-excitation of the even state into the waveguide, acting as a single collective state. The slower decay into the reservoir states only slightly modifies the details of the curve shape as compared to the single emitter case. This behavior is precisely the one predicted by the Markov approximation: both symmetric and antisymmetric qubit states act as a single emitter, decaying collectively into the waveguide with different decay rates that can be modified through the qubit-qubit separation.
 
 In figure \ref{FIGframes}(b) we move out of the Markovian regime, choosing the same parameters as in figure \ref{FIGgoingout}(b). In this case, a significant part of the wavepacket has been emitted into the waveguide modes before qubit 2 starts taking part in the system evolution. After the pulse reaches this second emitter, a typical interference pattern appears inside the qubit-qubit cavity. This irregular pattern lasts for a very short time, while photons escape in successive reflections. Last frame shows the case in which the transient regime is about to end and the collective one has arisen. This state will remain in that position as a stationary wave, decaying slowly into the reservoirs with a time scale $t \sim 1/\Gamma$.
 
 Finally, figure \ref{FIGframes}(c) shows the photonic probability density for the same parameters as in figure \ref{FIGultrastrong}, i.e., strong qubit-waveguide coupling. We can observe how the first emitter has fully decayed when the pulse reaches qubit 2. Successive reflections occcur when the pulse hits one of the emitters, but no collective state arises in this transient regime. The coupling is so strong that the interaction between a qubit and the guided photon finishes before the pulse reaches the other emitter, so that no interference pattern appears. Collective evolution of the qubits is thus shown to be supressed. All these frames are consistent with the discussion of figures \ref{FIGmarkovpops}, \ref{FIGgoingout} and \ref{FIGultrastrong}. They confirm that, in the single excitation subspace, an increase in the emitter-waveguide coupling supresses collective effects,  thus destroying the entanglement generation.
 
 As a final discussion, let us comment briefly on the feasibility of experimentally observing the described non-Markovian effects. As remarked above, non-Markovian evolution requires an increase in the qubit-waveguide coupling. However, a realistic, achievable coupling \cite{CitaCOupling} is typically non larger than $\gamma/\Omega \approx 0.01$. Although in this situation the dynamics deviates significantly from the Markovian predictions, for an easier observation the inter-qubit separation $d$ has to be increased [figure \ref{FIGgoingout}(b)]. This effectively increases the coupling as stated above, but brings out the problem of losses in the waveguide. For high $\beta$ factors, even a moderate separation $d \approx 10 \lambda$ is high when compared to the propagation length for both plasmonic waveguides \cite{GenerationSINGLEspp} and photonic crystal waveguides \cite{PCproplength}. On the other hand, extremely low losses in dielectric waveguides have been reported \cite{dielWGpropagationLENGTH}, but $\beta$ factors are usually much lower. A feasible observation of non-Markovian effects requires both a high $\beta$ and a high propagation length. Some systems have been already reported to achieve this goal \cite{BarangerNONmarkov}.

 \section{Conclusion}
 
 We have presented a full quantum electrodynamical solution to the qubit-qubit system coupled to a waveguide in the single excitation case. A complete set of eigenstates has been obtained for the first time, thus allowing us to study the population dynamics for every combination of parameters. Specifically, we have explored the validity of the Markovian approximation. Our results show that the Markovian results are recovered for low qubit-waveguide coupling. However, as this coupling is increased non Markovian effects start to appear. These deviations can be extremely large, so that all traces of Markovian dynamics are lost and the evolution of the system is completely different. In this high coupling regime, both qubits act independently and the entanglement formation is supressed. This is in opposition to what Markov approximation suggested, i.e., an improvement of the entanglement properties when the coupling is increased. Although the Markovian regime seems to be a better option for entanglement purposes, the new strongly coupled regime shows very interesting dynamics, as both qubit and photonic degrees of freedom play a key role in the evolution. As a consequence, this regime is a good framework for studying intense light-matter interaction phenomena, such as strong coupling in waveguide QED systems.
 
\section*{Acknowledgements}
 
 This work has been funded by the European Research Council (ERC-2011-AdG Proposal No. 290981) and by the Spanish MINECO under contract MAT2011-28581-C02-01.

\providecommand{\newblock}{}

\end{document}